\documentclass[reprint, amsmath, amssymb, aps,nofootinbib]{revtex4-1}
\usepackage{graphicx}
\usepackage{dcolumn}
\usepackage{bm}
\usepackage{bbold}
\usepackage{slashed}
\usepackage{soul}
\usepackage{times}
\usepackage{epstopdf}
\usepackage{epsfig}
\usepackage{graphicx}
\usepackage{xcolor}
\graphicspath{ {images/} }
\usepackage{amsmath}
\usepackage[titletoc, title]{appendix}
\usepackage[colorlinks,
            linkcolor=red,
            anchorcolor=blue,
            citecolor=green
            ]{hyperref}

\begin{document}
\title{Narrow exotic hadrons in the heavy quark limit of QCD}
\author{Yiming Cai}
\email{yiming@umd.edu}
\author{Thomas Cohen}
\email{cohen@physics.umd.edu}
\affiliation{Maryland Center for Fundamental Physics and the Department of Physics,\\
University of Maryland, College Park, MD, USA\\}

\date{\today}

\begin{abstract}
This paper focuses on tetraquarks containing two heavy quarks in the formal limit where the heavy quark masses are taken to be arbitrarily large. There are well-established model-independent arguments for the existence of deeply bound exotic $\bar{q}\bar{q}'QQ$ tetraquark in the formal limit of arbitrarily large heavy quark masses.  However, these previous arguments did not address the question whether tetraquark states exist close to the threshold for breaking up into two heavy mesons.   Such states are not stable under strong interactions---they can emit pions and decay to lower-lying tetraquark states.  This raises the issue of whether the nature of the heavy quark limit requires such states to exist as resonances that are narrow.  Here we present a model-independent argument that establishes the existence of parametrically narrow tetraquark states. The argument here is based on Born-Oppenheimer and semi-classical considerations.  The results derived here are only valid in a formal limit that is well outside the regime occurring for charm quarks in nature. However the analysis may give some insight in the experimentally observed narrow near-threshold tetraquark states containing a heavy quark and antiquark.
\end{abstract}

\maketitle

\section{Introduction}

There has been a renaissance in hadronic physics over the past decade and a half with the discovery of hadrons that seem to be quite clearly exotic as they contain a charm quark and antiquark in addition to light quark content \cite{guo2018hadronic,godfrey2008exotic}.  None of these are stable under strong interactions---they are resonances that can decay into a J/$\psi$ plus another hadron.  However these resonances are  narrow enough to be clearly observed.   Many of them are found quite close to threshold for the decay into two heavy hadrons.   The existence of such near-threshold narrow exotic states may well be related to the fact that the charm-anticharm pair is significantly heavier than typical hadronic scales.  One key question is the extent to which the narrowness of such states is driven by the heaviness of the charm quark.

This paper seeks to get some insight into that issue by studying a system that is in some ways analogous: tetraquark systems containing two heavy quarks (rather than a heavy quark and antiquark) in the artificial limit where the quark mass becomes arbitrarily large.  As will be shown in this paper, in that artificial limit a large number of tetraquarks with fixed quantum numbers emerge, including some near threshold.  These near threshold states, like those the charm-anticharm system, are unstable under strong interactions.  (The double-heavy tetraquarks can decay via the emission of pions into lighter tetraquarks).  The virtue of studying this analogous problem in this limit is that one has theoretical control of the problem to leading order in a systematic $1/m_Q$ expansion (where $Q$ indicates a heavy quark)  by using Born-Oppenheimer and semi-classical analysis along with known features of QCD.   With these methods, it is possible to demonstrate that these tetraquark states exist as resonances and that these are narrow:  the width goes to zero as $m_Q \rightarrow \infty$.  It remains an open question as to whether the techniques developed here can be extended to heavy tetraquarks containing a heavy quark and a heavy antiquark.


It is long been known that  clearly discernible quantum-number exotic hadrons exist in certain limits of QCD.  \footnote{For example, in the large number of colors ($N_c$) limit, hadronic resonance with the quantum numbers of hybrid mesons such as  $J^{PC}=1^{-+}$ must exist\cite{cohen1998quantum}; their widths can be shown to scale with $N_c$ as $1/N_c$.   This indicates that  these hadrons are narrow at large $N_c$ and, thus, clearly identifiable as distinct resonances.}  Demonstrations of this type based on limits of the theory are not definitive in describing the physical world, but they show clearly, that the structure of nonabelian gauge theories such as QCD does not forbid exotic hadrons.  In this spirit, the present paper probes the question of what the heavy quark limit of QCD tells us about exotic  hadrons containing two quarks.   

This is motivated in part  by the discovery of the so-called X, Y and Z mesons and subsequently the P baryons\cite{guo2018hadronic,godfrey2008exotic}.  The  various X, Y, Z and P resonances  from their masses and decays generically ``look like'' some form of state with $c \bar{c}$ pair plus light quark content.   While strictly speaking none of these have exotic  quantum numbers, in the realm of hadronic physics,  QCD dynamics suppresses the creation and annihilation of heavy quark pairs\cite{georgi1992heavy}.   Indeed, this is a prerequisite for the viability of NRQCD\cite{grinstein2000modern}.   Thus to good approximation one can think of QCD dynamics inside of hadrons as preserving the number of charm quarks and charm antiquarks separately and in that sense a number of these states have exotic quantum numbers.  It is worth recalling that while there is strong evidence for strong-interaction stable non-exotic doubly-charmed baryons\cite{olsen2017non}, as of now there are no known doubly-charmed or doubly-bottomed tetraquarks.  However, it is also worth noting that this does not mean that these states do not exist in QCD: it could indicate that they are very difficult to make in typical laboratory experiments.

In this paper, it is shown in a model-independent way that
\begin{itemize}
\item Exotic tetraquarks of the $QQ$ type must exist as narrow resonances in the formal  heavy quark limit in at least some spin-flavor channels.
\item For sufficiently heavy quarks,  multiple narrow exotic hadrons will exist with fixed quantum numbers.   As the heavy quark masses increase the number of such states grows without bound.
\item In channels with exotic hadrons,  exotics exist that are below, but parametrically close, to the threshold for decaying into two non-exotic hadrons each of which contains a heavy quark.
\end{itemize}

The analysis leading to this conclusion depends on an intricate interplay of various semi-classical and Born-Oppenheimer-type approximations. However, the underlying physical idea is straightforward:  As $m_Q \rightarrow \infty$,  a  scale separation develops between the slow scales associated with the motion of the heavy quarks and the motion at characteristic hadronic scales.   This forces a decoupling between a tetraquark state and processes at the the hadronic scale, including the pion emissions that lead to tetraquarks' width.  

For simplicity the analysis done here will assume that the light quarks in the state are either up or down (rather than strange) and that isospin is an exact symmetry.  However, these assumptions can be relaxed with no fundamental change in the argument.  Only strong interaction effects will be considered in this work: electro-weak interactions will be excluded and thus decays of heavy quarks into lighter ones will be ignored.

One subject that has received considerable attention is the extent to which various X, Y and Z states are ``really'' tetraquarks as opposed to  di-meson ``molecules'' \cite{chen2016hidden,chen2016hidden,guo2018hadronic}.   The perspective of this paper is that the question of  whether a  state has a ``molecular'' or ``tetraquark'' character is largely an issue of what type of description is appropriate in the context of models.   Since our intention is to learn something from QCD in a model-independent way , all of these  hadrons will be referred in this work as ``tetraquarks'',  without prejudice;  all of them contain (at least) two quarks and two antiquarks. 

There are  two significant  differences between tetraquarks containing $QQ$ and those with  $Q\bar{Q}$.    One  is the possible annihilation of the $Q \bar{Q}$ pair; the second is the possibility of a rearrangement in which the $Q \bar{Q}$ can form a color-singlet quarkonium state leaving behind an ordinary meson.  While  the annihilation of $Q \bar{Q}$ pairs is suppressed in the heavy quark regime, the second  issue is potentially serious.    The formalism developed in this work for $QQ$ tetraquarks does not account for possible rearrangement effects  in $Q\bar{Q}$ tetraquarks.  It is plausible that the formalism developed here could be modified to account for rearrangement and the conclusions  drawn here might also apply to these cases.  

There are well-established  and compelling arguments\cite{cohen2006doubly, manohar1993exotic, zouzou1986four}  for the existence of exotic  $\bar{q} \bar{q}' Q Q$ tetraquarks (where $q$ and $q'$ are light quarks with possibly distinct flavors).   This paper borrows  from one of these arguments\cite{manohar1993exotic}.   The key theoretical issue addressed in this paper goes well beyond what can be addressed directly  using the analysis of \cite{manohar1993exotic}.  In particular, in that work the tetraquark is treated as a non-relativistic bound-state of two heavy mesons, and the long distance potential in the Schr\"odinger equation is given by one-pion-exchange.  This set up is clearly sensible when the bound state is stable under strong interactions; below the pion-emission threshold such systems can be modeled as a two-body system. However such a description becomes problematic for near-threshold tetraquarks as the heavy quark limit is approached since these state can be shown to be unstable against the emission of pions.  The major theoretical advance in this work is the development of a consistent framework to treat such states to leading order in a $1/m_Q$ expansion.  The key result is that width of these states is suppressed in a systematic as $m_Q \rightarrow \infty$.

A word about scales: we are interested in how observables vary with the heavy quark mass and typically express this in terms of a dimensionless ratio with a generic  non-heavy hadronic scale denoted as $\Lambda$.  To render various expressions dimensionful we use explicit factors of $\Lambda$, which are taken to be an arbitrary but fixed value at the hadronic scale.  Part of the analysis involves long-ranged forces characterized by a length scale of $1/m_{\pi}$ and the dimensionless ratio of $m_\pi/m_Q$ also enters the problem.  In the light quark sector one often considers there to be a scale separation between $m_\pi$ and $\Lambda$;  such a separation is the basis for chiral perturbation theory\cite{scherer2003introduction}.  Here we are interested in asymptotic scaling as the heavy quark mass grows and it is convenient to consider $m_{\pi}$ to be of order $\Lambda$ but simply numerically small.  

In this paper, we use $m_Q$ to represent the mass of heavy quark, and $M_Q$ to represent the mass of a meson containing one heavy quark. Note that, the difference between $m_Q$ and $M_Q$ is of  relative order $\Lambda/m_Q$, which is suppressed by $m_Q$ and is neglected in our leading-order analysis. 

The basic approach depends on the fact that the distance between the heavy quarks changes slowly in the heavy quark limit; thus, for a certain range of kinematic variables, QCD for states with two heavy quarks becomes an effective two-body quantum mechanical system with this distance as the degree of freedom.  At very short distance, the color coulomb interaction between the heavy quarks will dominate the effective interaction; at intermediate distance, the form of the interactions is complicated.  This paper largely focuses on long distances.  At large distance between the heavy quarks, the lowest lying states of QCD in the heavy quark limit will, with arbitrarily good accuracy, be described by two heavy mesons.   At the longest distances, the interaction between these heavy mesons will be described with arbitrarily good accuracy via one-pion exchange.  Thus the QCD dynamics of these states at long distances, is well described as one-pion-exchange between two heavy mesons.

Our analysis connects QCD dynamics with hadronic dynamics when the separation between the heavy quarks is large, but the analysis is in fact model-independent. In particular, at no point does it rely on the constituent quark model.  This is important since intuition based on a constituent quark model can be misleading when the heavy quarks are separated by a large distance.

This paper begins with a brief review of the known salient features of tetraquarks containing a $Q {Q}$ pair.  The next section focuses on an analysis that shows in a simple treatment based on the Schr\"odinger equation as the heavy quark mass increases, multiple exotic states exist as bound states for each channel containing exotics and that most of these states are   unstable against strong interactions.  Following this, is the principal theoretical development of the paper: it will be shown that the full QCD Hamiltonian in the sector with exotic quantum numbers can be decomposed to a piece described  via the Schr\"odinger equation and a remainder, and that the effect of the remainder is to convert the  exotic bound states  of the Schr\"odinger equation into narrow resonances.   Next there is a brief section demonstrating that some of these  narrow resonances are close to the threshold for breaking up into two heavy mesons.  Following this is a discussion section that summarizes the result and explores the challenges posed by rearrangement effects and possible phenomenological implications of the results for tetraquarks containing charm-anticharm pairs.

\section{On the existence of tetraquarks with two heavy quarks \label{QQ}}
There have been two distinct arguments for the existence of strong-interaction  stable $\bar{q} \bar{q}' Q Q$ tetraquarks. 
However, we are interested in the near threshold states, which are not treated in the previous literature. Before stating our arguments, let us at first review the previous related works. The first is based on the attractive Coulomb potential between two heavy quarks in a $\bar{3}$ color configuration\cite{cohen2006doubly, manohar1993exotic}.  In the extreme heavy quark limit, this will lead to a  deeply bound diquark configuration, with a binding energy of order $\alpha_s^2 m_Q$, and a characteristic size of order $(m_Q \alpha_s)^{-1}$ where $\alpha_s$ is taken at the scale of the inverse size.  In the extreme heavy quark limit, there will exist a tightly bound diquark of small size in the $\bar{3}$ color configuration.  Because it is heavy, tightly bound and small  it will act   as a nearly point-like source of a color Coulomb field.  A single heavy antiquark  in the $m_Q \rightarrow \infty$ limit and viewed from the rest frame of the hadron  also acts  as a static source of a color Coulomb field in the $\bar{3}$ representation (its spin becomes irrelevant as $m_Q  \rightarrow \infty$ and the quark's recoil is suppressed).   Thus, the tightly bound diquark dynamically acts in the same way as a heavy antiquark.  A recent lattice simulation also supports this theoretical prediction\cite{francis2017lattice}.

This doubly-heavy-diquark-antiquark (DHDA)  duality means that so long as the system is well below the excitation energy of the heavy diquark---formally of order $\alpha_s^2 m_Q$ and thus well above $\Lambda$---every hadronic state in the system with two heavy quarks has an analog state for the system with a heavy antiquark (strictly speaking with a set of states with nearly degenerate masses which differ by the spin of the heavy antiquark).   The DHDA duality implies that since antibaryons containing heavy antiquarks exist  as strong-interaction stable states  in QCD, then so too will  $\bar{q} \bar{q}' Q Q$ tetraquarks, at least in the extreme heavy quark limit.  

In practice,  the charm mass is significantly too small for the DHDA duality to apply; the bottom mass is at best quite marginal\cite{cohen2006doubly, manohar1993exotic}.  However, on a theoretical level the argument remains valid for the case of extreme masses.  Unfortunately, this argument  provides little insight into near threshold states.

There is another argument\cite{cohen2006doubly, manohar1993exotic}  that also requires strong-interaction stable $\bar{q} \bar{q}' Q Q$  tetraquarks.  The argument treats strong-interaction stable tetraquarks of the  $\bar{q} \bar{q}' Q Q$  type as a bound state problem of two interacting heavy mesons, each one containing a heavy quark $Q$.  At first blush, this may seem as though it is valid only if  the ``molecule'' picture is correct.  In fact, however, it holds regardless of the underlying structure of the tetraquark state.   If the processes under consideration restricts the motion so that no additional light particles are created in the dynamics and also ensure that the heavy mesons remain distinct, that they move non-relativistically and that they remain at large distances from each other,  then the system can be described accurately by a  Schr\"odinger equation.

Consider what happens in general to a system containing two subsystems whose relative dynamics is describable by Schr\"odinger equation if some parameter of the problem, $\lambda$, is adjustable and this allow for the mass of the two subsystems to increase (to arbitrarily large size) while only weakly affecting the potential.  For simplicity, let us first focus on an s-wave channel without mixing with purely non-relativistic kinematics.  The time-independent Schr\"odinger equation for the $n^{\rm th}$ bound state, given in terms of the radial coordinate is then given by 
\begin{equation}
\left(-\frac{1}{2 \mu(\lambda)} \frac{\partial^2 }{d r^2} +V(r;\lambda) \right ) u_n (r; \lambda) =E_n(\lambda)  u (r; \lambda)
\end{equation}
where $\mu \equiv  m_1 m_2/(m_1+m_2)$ is the reduced mass,  $\lambda$ dependence of all quantities  is made explicit, $u$ is normalized so that $\int dr |u(r)|^2=1$ and the potential goes to zero at infinity.    It should be clear that increasing $\mu$ plays the same role in determining the existence of normalizable eigenstates as increasing $V$, since only the product $\mu V$ matters.  Thus, if $V$ is attractive then at large enough $\mu$, the system will have bound states.

More formally, suppose that the potential  remains finite and  well-defined in the large $\lambda$ limit while the reduced mass asymptotes to a linear dependence:
\begin{equation}
\lim_{\lambda \rightarrow \infty} V(r;\lambda) =V_0(r) \; \; , \; \;
\lim_{\lambda \rightarrow \infty} \frac{\mu(\lambda)}{\lambda}  = \tilde{\mu} \; 
\end{equation}
 and moreover there is with at least some region in $r$ for which the  $V_0(r) < 0$  ({\it i.e.} the potential has an attractive region).
In this situation, it is trivial to show via a variational argument that  at sufficiently large $\lambda$ bound states must exist.  \footnote{ It is always possible to  construct a class of $\lambda$-dependent normalized trial wave functions $u_{\rm trial}(x,\lambda)$ with the properties that  $\lim_{\lambda\rightarrow \infty} |u_{\rm trial}(r,\lambda)|^2=\delta(r-r_0)$ and   $\lim_{\lambda\rightarrow \infty} \frac{|u'_{\rm trial}(r,\lambda)|^2}{\lambda}=0$, from which it follows that
$$\lim_{\lambda\rightarrow \infty} \langle u_{\rm trial} |H|  u_{\rm trial}\rangle= V(r_0) < 0 \; .$$
Since the energy of the true ground state  of the system needs to be below that of the trial wave function, it follows that the ground state has negative energy--{\it i.e.} that a bound state exists, provided that there is not an additional open channel with lower energy than the variational state.

 A key point is that the description in terms of a simple one-degree of freedom Schr\"odinger equation need not be valid for the entire Hilbert space,  provided that one can choose a trial wave-function  with negative expectation value of energy in the domain where the description {\it is} valid.  Then the ground state of the system must have negative energy, even if it is outside the domain of validity of the Schr\"odinger   dynamics.  }  An analogous argument holds for any partial wave and to systems with mixing of degrees of freedom (such as spin and orbital, or distinct flavors) and to systems with relativistic kinematics.   
 
  The previous argument should hold for  $\bar{q} \bar{q}' Q Q$  tetraquarks, since the lowest threshold for fixed quantum numbers will be for breaking into two heavy hadrons.   The long distance part of the strong-interaction potential  in the heavy quark limit is well known.  It is given by one-pion exchange.  The precise form that the long-distance potential takes depends on heavy quark symmetry, in particular the fact the spin of the light decouples from the dynamics at large heavy quark mass.   Heavy-hadron chiral perturbation theory (HH$\chi$PT) \cite{cho1993heavy}  gives the correct form\cite{manohar1993exotic}.  However, the form depends only he fact that the pion is the lightest hadron and does not depend on the validity of chiral perturbation theory.
  
Recall that the pseudoscalar and vector heavy mesons form a multiplet that becomes degenerate in the heavy quark limit; the spin of the heavy quark becomes irrelevant in each meson.   We denote these two generically as $H$ and $H^*$ and assume that the system is sufficiently  close to the heavy quark limit for  $H$ and $H^*$ to be treated as degenerate. It is clear that because the pion is a pseudoscalar, it couples to an angular momentum and because the spin of heavy quark is irrelevant it couples to the internal angular moment of the light degrees of freedom in the heavy meson. 


Since the heavy quark spin, $\vec{S}_Q$  decouples from the dynamics it is useful to decompose the total angular momentum in the rest frame, {\it i.e.} the tetraquark spin , $\vec{S}$  as
\begin{equation}
\vec{S} = \vec{S}_Q + \vec{J} \; \; {\rm with} \; \;  \vec{J} =\vec{L}+\vec{S}_l
\end{equation}
where $\vec{S}_l$ is the total spin of the light degrees of freedom and $\vec{L}$ is the orbital angular momentum.   $\vec{J}$ and $\vec{S}_Q $ are separately conserved.
Many possible spin-isospin channels exist for this system and the one-pion exchange potential depends on the channel; the allowable channels are constrained by Bose symmetry between  the  two heavy mesons.  We fix the $J$, $I$ and parity quantum numbers for the analysis  and can subsequently couple these with $ \vec{S}_Q $ to obtain the allowed physical states.

Our initial focus is on  states with $J=0$ and positive parity.   Such states automatically have $S_l=0$ and spatial states with  $L=0$ \footnote{Note that, the total spin of the light degrees of freedom has been chosen to be 0, but this is only part of the spin of the tetraquark. The allowable spin of the heavy quarks has decoupled from the dynamics, and can always be chosen to be in any configuration consistent with the Pauli principle. This may affect the possible spins of the final tetraquark state.}.  This channel is particularly simple to analyze as the long-range potential is central and does not mix orbital partial waves.    For $S_l=0$ states one-pion exchange  is given by a simple Yukawa potential:
 \begin{equation}
 \tilde{V}_{\rm long}(r) = -\left ( I (I+1) - \frac{3}{4}  \right ) \frac{g^2 m_\pi^2}{\pi f_\pi^2} \frac{e^{- m_\pi r}}{r}
\label{lrp}  \end{equation}
where $I$ is the total isospin of the state, $g$ is the coupling constant for  $H-H^*-\pi$ coupling (evaluated at $q = 0$) and $f_\pi \approx 93$ MeV is the pion decay constant; the quantities $g$, $f_\pi$ and $m_\pi$ are all understood to be at their heavy-quark limit.  An obvious and significant point is that for states with $I=1$ the potential is negative, thereby ensuring the existence of a tetraquark in the $m_Q \rightarrow \infty$ limit.  Moreover, $I=1$ ensures that the system has exotic quantum numbers.  Note that the essential role played by the one-pion exchange potential in this argument was to ensure a long-distance attractive channel.

This is sufficient to demonstrate that tetraquarks of the $\bar{q} \bar{q}' Q Q$  exist as bound states in the heavy quark limit; this argument holds independent of the argument based on the DHDA duality and independently of whether the ``molecule''
 picture is correct.   \footnote{If one were ignorant about whether a deeply bound doubly heavy tetraquark existed, there are two possibilities: deeply bound tetraquarks exist and decays of weakly-bound tetraquarks into one via pion emission are possible.  This could invalidate the potential-based analysis above---but requires a tetraquark to exist.    If  deeply bound states did not exist, then the potential based  argument is valid and it predicts that a bound tetraquark must exist in the heavy quark limit.  In either case, a tetraquark exists.}  One might be concerned about the form the potential, since strictly speaking, the argument as given only applies to the particular quantum numbers ($I=1$, $J=0$, positive parity),  however analogous arguments exist for any channel that is attractive at long distance, which is our main focus.

The argument does not go through for the case of tetraquarks with $Q \overline{Q}$ content.  The argument that there  are  states with negative energy relative to the two heavy meson threshold remains valid.  However there are states below that threshold that do not correspond to tetraquarks, for example non-resonant states of a pion plus a $J/\psi$.  Of course, if there is a dynamical reason this open channel is weakly coupled to the dynamics under consideration it is plausible that arguments analogous to the ones here might imply narrow resonant states.

\section{On the existence of multiple tetraquarks with fixed quantum numbers \label{Many}}

Having reviewed related work, we will begin a discussion with a toy problem, a two body Schr\"odinger equation. We do this because, as noted briefly in the introduction, in the heavy quark limit, states of QCD, containing two heavy quarks, are arbitrarily well-described at large quark separations, by two heavy mesons interacting via potential. Moreover it is known to be a Yukawa type potential at large separations. 


In this section,  it is shown that a Schr\"odinger equation with a  potential whose long-range part  is given in  Eq.~(\ref{lrp}) supports a  large number of distinct bound states with fixed quantum numbers in the heavy quark limit.  
Superficially this seems to imply that at very large heavy quark masses there are parametrically many distinct exotic hadrons containing two heavy quarks. 
This turns out to be correct, but the analysis is somewhat subtle: most of these states 
do not exist as bound states describable in terms of a  two-body Schr\"odinger equation  but as  narrow resonances in a description with more asymptotic degrees of freedom.   Never-the-less a Schr\"odinger equation description is a good starting point. This issue will be addressed in the following section. 

\subsection{Description based on the two-body  Schr\"odinger equation \label{A}}

A semi-classical analysis explains why the  Schr\"odinger equation description  implies a large number of bound states at large heavy quark mass. Such an analysis is far simpler in contexts where there is only a single dynamical degree of freedom.  For this reason we start  with a description of $I=1$, $J=0$, positive parity states; these do not involve coupled channels.  States with other quantum numbers involve a slightly more complicated analysis (addressed in the next subsection) but the conclusions are qualitatively the same.

  For  systems with a single dynamical degree of freedom, standard semi-classical analysis implies that the number of bound states between energy 
 $E_0$ and $E_0+\Delta E$ is well-approximated by
\begin{equation} \begin{split}&N(E_0,\Delta E) \approx \frac{ \int_{E_0}^{E_0+\Delta E} \, {\rm d} E  \int{ \rm d}  p  \, {\rm d} q \,  \delta(E-H(p,q) )}{2 \pi} \\ & = \frac{   \int{ \rm d}  p  \, {\rm d} q \,  \Theta (E_0 +\Delta E-H(p,q) ) \Theta (H(p,q) -E_0)}{2 \pi} \label{nofE}\end{split} \end{equation}
where $H$ is the classical Hamiltonian and $\Theta$ is a Heaviside step function.  The approximation becomes increasingly accurate as $ N(E_0,\Delta E)$ becomes large.  Suppose that the Hamiltonian is of the form $H=\frac{p^2}{2 \mu} + V(q)$ where $\mu$ is the reduced mass and $V(q)$ is independent of $\mu$. It is trivial to see that phase space area between $E_0$ and $E_0 +\Delta E$ is directly proportional to $\sqrt \mu$ and accordingly so is $ N(E_0,\Delta E)$.

Let us first apply this to any arbitrary Hamiltonian for an $L=0$ channel with a single degree of freedom whose the long-range potential is given in Eq.~(\ref{lrp}) and whose kinetic  is given by $p^2/(2 \mu)$.  Denote $N^{\rm long}(E_0,\Delta E)$, as the contribution to integral in Eq.~(\ref{nofE}) coming from values of $r$ that are well described by the long-distance potential.  Provided that  as $N^{\rm long}(E_0,\Delta E) \gg 1$ in some energy range, the system is in the semi-classical domain.  If we identify this problem with the physics problem of interest (the long distance  potential associated with the $I=1$, $J=0$, positive parity tetraquark channel),  $\mu=M_Q/2$.  Thus,  as $M_Q \rightarrow \infty$, $N^{\rm long}(E_0,\Delta E)$ diverges as $\sqrt{M_Q}$; at large $M_Q$, the total number of  bound doubly heavy tetraquark states  in any energy grows as  $\sqrt{M_Q}$:   at large $M_Q$, there a large number of tetraquarks. \footnote{In a simple quark model type of approach\cite{lee2009stable}, it has been argued that on the basis of a chromomagntic interaction, the most attractive channel is $I = 0$. However, our purpose in this analysis is not to find the most bound tetraquark, bu  rather to study states with exotic quantum numbers—hence our focus on the exotic $I=1$ channel.  As noted in Sec.II, in the heavy quark limit, the existence of any attractive region for a given set of quantum numbers implies a tetraquark state.  Recall that, our principal interest is for the long distance region for which the Yukawa interaction ensures that there exists an attractive $I=1$ channel. For our purpose, this is sufficient. In any case, in this region the chromomagnetic interaction is negligible; it is a short distance effect. Whether or not the short distance potential for I=1 is the most favorable energetically is not relevant given that the long distance potential is attractive.}


The 
total number of bound doubly heavy pentaquark states, $n_{\text{tot}}$ is then bounded by 
\begin{equation} 
n_{\text{tot}} \ge n_B \equiv N^{\rm long}(-B,B) \ge \int_{r_B}^\infty {\rm d} r \, \frac{ \sqrt {-M_Q   \tilde{V}_{\rm long}(r) }}{ \pi} .  \label{ineq}\end{equation} 
$N^{\text{long}}$ is the total number of semi-classical bound states with energy less than $\tilde{V}_{\rm long}(r_B)$, where $r_B$ is sufficiently large so that $\tilde{V}_{\rm long}(r_B)$ closely approximates the potential; $r_B $ is defined implicitly through $\tilde{V}_{\rm long}(r_B)=-B$, and $n_B$ is the number of bound states with binding energy less than $B$. 
The second inequality in Eq.~(\ref{ineq}) comes from evaluating the integral only over the region greater than $r_B$ rather than the entire long distance region.  Finally, using the explicit form for  $\tilde{V}_{\rm long}$, it is straightforward to prove that at sufficiently large $\mu$
\begin{equation} n_B >\frac{\sqrt{M_Q B}}{2\sqrt{2}    \, m_\pi} \sim  \frac{\sqrt{M_Q B}}{ \Lambda} \; .\label{ineql3} \end{equation}
where we take $m_\pi$ to be of order of a characteristic hadronic scale.  The analysis is valid provided that  $n_B \gg 1$.
If $B$ is held fixed, $n_B$ the number of bound states with binding energy less than $B$, grows  with $M_Q$ at least as fast as  $ \sqrt{M_Q}$.

Unfortunately this analysis based on a  Schr\"odinger equation  need not imply QCD in the heavy quark limit also has numerous bound tetraquark states.   QCD allows for the the emission of mesons which is beyond the regime of validity of the potential model description.   However, as will be discussed in Sect.~\ref{MCD}, a remnant of these would-be bound states survive in the heavy mass limit of QCD but as narrow resonances rather than bound states.

\subsection{Other quantum numbers \label{OQN}}

The argument of the previous subsection can be generalized to attractive channels with quantum numbers in which the tensor force in the one-pion-exchange mixes the $L$ quantum number;  this leads to a coupled-channel problem.  Fortunately, the kinematics of the problem allows one to choose a basis in which the channels decouple up to corrections that vanish in the heavy quark mass limit--reducing the problem to an effective single degree of freedom.  The argument is in the spirit of the Born-Oppenheimer approximation\cite{born1927quantentheorie}, which been used extensively  in the context of heavy quark exotic states such as heavy-quark hybrids\cite{berwein2015quarkonium, brambilla2019spin,brambilla2018born}.)

In the heavy quark limit, the two mesons move slowly.  The potential that couples the various channels is a matrix (which depends on the separation between the mesons, $r$), with diagonal terms within a channel and off-diagonal terms for cross-channel coupling.   For fixed $r$ the potential matrix can be diagonalized; the eigenvalues give the value of the potential for the ``new'' channels and the eigenvectors give these new channels in terms of the old ones.  Since the mesons move slowly, the variation is adiabatic.  If the system is the lowest eigenstate of the potential for one value of $r$, it will tend to remain in the lowest eigenvalue for all values of $r$.  This tendency  becomes perfect as $M_Q \rightarrow \infty$;  the system acts like a single channel with potential given by the lowest eigenvalue.

To make this explicit, consider the general form of the Schr\"odinger equation for a  potential problem with $k$ channels:
\begin{equation}
\begin{aligned}
& \left ( -\frac{\overleftrightarrow{1} }{2 \mu} \partial_r^2  +  \overleftrightarrow{V} \right ) \vec{\psi} = E  \vec{\psi}  \; \;   {\rm with}  \; \; \vec{\psi}(r) \equiv \left (\begin{array}{c}   \psi_1(r)\\  \psi_2(r)\\ \vdots \\ \psi_k(r) \end{array}  \right )\\
& {\rm and} \; \; \; \;   \overleftrightarrow{V}(r) \equiv \left (\begin{array}{cccc}   V_{11}(r)& V_{12}(r) &...&V_{1k}(r)\\   V_{21}(r)& V_{22} (r)&...&V_{2k} (r)\\
  \vdots & \vdots & \vdots &\vdots\\  V_{k1}(r)& V_{k2} (r)&...&V_{kk}(r)  \end{array}  \right ) \; ,
\end{aligned} 
\end{equation} where $\mu$ is the reduced mass.  
$ \overleftrightarrow{V}(r)$ has a magnitude of order $\Lambda$ and varies over a distance of order $\frac{1}{\Lambda}$; while the kinetic term is controlled by $\mu \sim M_Q$.

$ \overleftrightarrow{U}(r) $ diagonalizes $\overleftrightarrow{V}(r)$:  $\overleftrightarrow{\tilde{V}}(r) \equiv   \overleftrightarrow{U}(r)\overleftrightarrow{V}(r)  \overleftrightarrow{U}^\dagger(r)$ is diagonal.    Defining $\vec{\tilde{\psi}} (r) \equiv  \overleftrightarrow{U}(r) \vec{\psi}(r)$ yields
  \begin{equation}
  \begin{aligned}
&   \left ( H_0 + H_1 +H_2   \right ) \vec{\tilde{\psi}} = E  \vec{\tilde{\psi}} \; \; {\rm with} \\
& H_0=\frac{-\overleftrightarrow{1} }{2 \mu} \partial_r^2  +  \overleftrightarrow{\tilde{V}}(r) \; \;  , \; \; H_1 = \frac{1}{\mu}   \overleftrightarrow{U}'(r) \overleftrightarrow{U}^\dagger(r) \partial_r  \\
&{\rm and} \; \;  H_2 = - \frac{1}{2\mu}   \overleftrightarrow{U}''(r) \overleftrightarrow{U}^\dagger(r) \;. 
\end{aligned}  \end{equation}

Returning to the tetraquark problem, $H_0$ is diagonal.  Moreover,  since spatial derivatives of $ \psi$ for the  semi-classical bound states are of order $\sqrt{M_Q \Lambda}$, nominally  $H_0 \sim \Lambda$,  $H_1 \sim \sqrt{\frac{\Lambda^{3}}{M_Q}}$ and $H_2 \sim \frac{\Lambda^2}{M_Q}$.  The $H_0$ term dominates at large $M_Q$; $H_1$ and $H_2$ can be treated as perturbations  $H_1$ is entirely off-diagonal and only contributes to the energy at second order.  Thus, the leading contributions to the energy from both $H_1$ and $H_2$  are of order $\frac{\Lambda^2}{M_Q}$. 

 In the reminder of this paper we will be dropping all effects of order  $\frac{\Lambda^2}{M_Q}$.   At this order $H_1$ and $H_2$ are negligible.    The diagonal nature of $H_0$ implies the lowest eigenvalue of $\overleftrightarrow{V}(r)$ acts as a potential for a problem with a single radial degree of freedom.   Once this fact is recognized, one can immediately exploit the arguments of  Subsection \ref{A} without further work for all quantum numbers with angular momentum of order unity and attractive interactions at long distances.  As $m_Q \rightarrow \infty$, there are multiple bound tetraquark states in all attractive channels.  
 
In the case of a purely s-wave interactions, there is no long-range repulsion due to a centrifugal barrier.  For $L \ne 0$ this is not the case.   Since the analysis involves the nature of the long-distance interaction, this might seem problematic.  However, provided that $J$ is  of order unity ({\it i.e.} $L \sim \left (\frac{M_Q}{\Lambda} \right )^0$) , then the orbital angular momentum, $L$, will be as well and the centrifugal term is negligible.

\section{A more complete description \label{MCD}}

The DHDA duality argument  of Sect.~\ref{QQ} implies the existence of tetraquarks, with  a binding of order $\alpha_s^2 m_Q$ relative to the threshold for dissociation into two heavy mesons---at least in the extreme heavy quark limit.  The energy of these states are  below that of the putative  tetraquarks discussed in the previous subsection (which generically have binding of order $\Lambda$) by more than $m_\pi$.  Moreover there is no symmetry that prevents a decay of such tetraquark states via the emission of one or more pions.  Given the totalitarian principle\cite{gell1956interpretation} of particle physics---that which is not forbidden is mandatory---weakly bound tetraquark states will decay via pion emission.  An analysis that does not this is incomplete.

While the potential-based argument is incomplete for bound tetraquarks states of order $\Lambda$ below threshold, the argument may still be valid basis for describing long-lived but none-the-less unstable tetraquark  resonances.    

An analogy can be made to the hydrogen spectrum: a coulomb potential description gives rise to a set of stable discrete levels.  However,  QED allows for the emission of photons:  none of the excited states are stable.  Never-the-less, the energy levels found in the Schr\"odinger  equation with a coulomb potential provide a very useful description: the states are sufficiently long lived that their decay rates---the widths---are much smaller than the energy differences between  a state and a neighboring state.    Our goal in this section is to show from QCD that in the heavy quark limit,  the decay rates of the near-threshold tetraquarks  are  small compared to the level spacing; these states  exist, but  as  narrow tetraquark resonances as opposed to bound states.  

At early stages in the analysis effects of relative order $\Lambda/m_Q$ are dropped.   Thus, effects with energies of order $\Lambda^2/m_Q$ are neglected.   For large $m_Q$, the neglect of such effects will not generically affect the existence of particular tetraquarks states including near-threshold states. 

For simplicity, the analysis done here will be for the attractive positive parity $I=1$, $J=0$ channel considered earlier that avoids mixing of partial waves.  However, as noted on Subsection \ref{OQN}, channel mixing effects are of relative order $\Lambda/m_Q$ for other channels and hence can be neglected to the order at which we are working; and thus the analysis goes over to all attractive channels without substantial change.

In the analysis  we show that generic tetraquark states of order $\Lambda$ below threshold have widths, $\Gamma$  and level spacing $\Delta E$ that scale with $m_Q$ as,
\begin{equation} 
 \Gamma  \sim \frac{\Lambda^2}{m_Q} \;  \;   , \; \; \Delta E \sim \sqrt{\frac{\Lambda^3}{m_Q}}\ \; . \label{GammaScal}  
 \end{equation}
 \begin{equation}
\frac{\Gamma}{\Delta E}  \sim \sqrt{\frac{\Lambda}{m_Q}} \; .  \label{narrow}
\end{equation}
$\frac{\Gamma}{\Delta E}$ goes to zero as $m_Q$ goes to infinity; the resonances are much narrower than their separation.  This is sufficient to establish that the tetraquarks are parametrically narrow.   

It is straightforward to see  from  standard semiclassical analysis  why $\Delta E$ scales as in Eq.~(\ref{GammaScal}).  Our principal goal in the remainder of this section is to demonstrate the scaling of $\Gamma$  in  Eq.~(\ref{GammaScal}).  The demonstration is based on a Born-Oppenheimer-like\cite{born1927quantentheorie} separation of the motion of the fast degrees of freedom (light quarks, gluons) from the slow motion of the heavy quarks.  In this respect the approach is similar in spirit  to pNRQCD\cite{brambilla2000potential}.    The characteristic velocity of the heavy quarks in the state is of order $\sqrt{\frac{\Lambda}{m_Q}}$ corresponding to a three momentum $\sim \sqrt{m_Q \Lambda}$ while the characteristic momentum scale of the light quarks and gluons in the state is $\Lambda$.  

Since the states in question are initially described as non-relativistic bound states, it is natural to use a Hamiltonian description;  for simplicity we will use the non-relativistic convention of quantum states normalized to unity rather than covariant normalization.   Moreover, it is convenient to work in a very large but finite rectangular box with periodic boundary conditions.  This ensures that the center of mass momentum remains a good quantum number, but is quantized to discrete values allowing normalizable states.  By working in the space of $\vec{P}_{\rm cm} = 0$ states, we decouple center of mass motion.

The approach begins with the leading order  NRQCD Lagrangian\cite{bodwin1995rigorous} for the light degrees of freedom coupled to nonrelativistic heavy quarks.  At leading nontrivial order, the number of heavy quarks is held fixed since pair creation is suppressed.  Since we are interested in the sector with two heavy quarks (which we will take to be the same species), we can write an effective Lagrangian containing exactly two heavy quarks:
\begin{equation}
 {\cal L}= Q^\dagger \left (-i D_0  - \frac{\vec{D}^2}{2 m_Q} \right) Q  +   {\cal L}_{\rm fast} 
\label{NRLag}\end{equation}
where $Q$ is the nonrelativistic heavy quark field, $D$ represents a covariant derivative and  ${\cal L}_{\rm fast}$ is the Lagrangian density for the fast degrees of freedom---the gluons and light quarks.   This is a valid representation of the full theory in the extremely heavy quark limit.  The next step is to choose to work in the  Coulomb gauge \cite{christ1980nh,schwinger1962non,szczepaniak2001coulomb}.   The final result will not depend on the choice of gauge. 

 One can  reexpress the physics in terms of the QCD Coulomb-gauge Hamiltonian, $H$.  First we introduce a projection operator $\hat{{\cal P}}^{\rm phys}$ which projects states from the full Hilbert space onto physical states consistent with the gauge condition and Gauss's law.  Such physical states are, of course, global color singlets.  We define a physical Hamiltonian: $\hat{H}^{\rm phys}=\hat{{\cal P}}^{\rm phys} \hat{H} \hat{{\cal P}}^{\rm phys}$ which acts in the Hilbert space of physical states.

 $\hat{{\cal P}}^{\rm phys}$ is also chosen to project on to a particular class of physical states: $\vec{P}_{\rm cm}=0$ states with the quantum number of interest.   For simplicity here we will first consider states containing two identical  heavy quarks with $I=1$, $J=0$ and positive parity.  The physical Hamiltonian for the system with fixed quantum numbers is $\hat{H}^{\rm phys}=\hat{{\cal P}}^{\rm phys} \hat{H} \hat{{\cal P}}^{\rm phys}$.  The time-independent physical eigenstates $|\phi\rangle$ of the system can be cast in the form
\begin{equation}
\hat{H}^{\rm phys} |\phi\rangle =E_\phi  |\phi\rangle \label{eigen1}
\end{equation}
 While explicitly constructing and computing with $\hat{H}^{\rm phys}$ and its eigenstates is not  practical,  one does not need to  compute with it  explicitly to deduce key scaling properties.

The next step is in the spirit of the Feshbach projection operators\cite{feshbach1958unified,feshbach1962h}.  We start by assuming that there exists a  projection operator that, when acting on the physical Hamiltonian, reduces the description of  one of nonrelativistic quantum mechanics with a single spatial degree of freedom $r$ (corresponding to the separation of  the heavy quarks) and a local potential that depends on $r$  whose long distance behavior is given  in Eq.~(\ref{lrp})).  (More precisely, it  yields such a description up to corrections of relative order $\Lambda/m_Q$).     Let us denote such a projector as $\hat{{\cal P}}^{\rm r}$; its complement is $\hat{{\cal Q}}^{\rm r}=(1-\hat{{\cal P}}^{\rm r})$.   By construction, the space of states onto which  $\hat{{\cal P}}^{\rm r}$ projects are tetraquarks with no additional pions.   We require that $\hat{{\cal P}}^{\rm r}$ commutes with $\hat{{\cal P}}^{\rm  phys}$.    In fact, as will be shown later in this section, such a projection operator exists.  For now, let us assume it to be so and follow the consequences.


The full physical Hamiltonian can be broken into two parts:
 \begin{equation}
 \begin{split}
 &   \hat{H}^{\rm phys } = \hat{H}_0 + \hat{H_I }\;  \; {\rm with} \\   &  \hat{H}_0 \equiv \hat{{\cal P}}^{\rm r}  \hat{H}^{\rm phys } \hat{{\cal P}}^{\rm r}  +  \hat{{\cal Q}}^{\rm r}  \hat{H}^{\rm phys } \hat{{\cal Q}}^{\rm r} \\   &H_I \equiv \hat{{\cal P}}^{\rm r}  \hat{H}^{\rm phys } \hat{{\cal Q}}^{\rm r}  +  \hat{{\cal Q}}^{\rm r}   \hat{H}^{\rm phys } \hat{{\cal P}}^{\rm r} \;  ;
  \end{split}
\label{PQ2} \end{equation}
we assume that there is a meaningful sense in which $  \hat{H}_0$ is the dominant term.  We can verify this assumption {\it a posteriori}.

By hypothesis $ \hat{H}_0$  acting on the space of states $|\phi \rangle_r$,  behaves like a local one-dimensional potential problem in a relative coordinate  with a potential that at long distance is given in Eq.~(\ref{lrp}).   Let us denote a typical state   in the $j^{\rm th}$ state abstractly as  
\begin{equation}
\hat{H}_0 |\phi_j\rangle_r=   
\left (E_j+2 M _Q+{\cal O}\left(\frac{\Lambda^2}{M_Q} \right)\right ) |\phi_j\rangle_r
\end{equation}
where $E_j$ is defined relative to the threshold for dissociation into two heavy mesons.  It is useful to define our effective two-body  Hamiltonian with the constant $2 M _Q$ removed
$\hat{H}_0^{\rm 2 \, body} \equiv \hat{H}_0 - 2 M_Q$  so that up to corrections of relative order $\frac{\Lambda^2}{M_Q}$, which will be neglected,  $\hat{H}_0^{\rm 2 \, body} |\phi_j\rangle_r= E_j | \phi_j\rangle_r$.  


Associated with 
 $|\phi_j\rangle_r$  are position space wave functions $\phi_n(r)$ satisfying 
\begin{equation}
 \begin{split}
& \left(\frac{-\partial^2_r}{M_Q} +V(r) \right)\phi_n(r)=E_n \phi_n(r) \\ &{\rm with} \; \;  \int_0^\infty dr \,  |\phi_n(r)|^2=1 \; .
\end{split} 
\label{SE}\end{equation}   
The (reduced) mass in the kinetic term is $M_Q/2$, half the heavy meson mass.  In fact, the kinetic term associated with the heavy quark mass has an $m_Q$ rather than $M_Q$ but the difference in using   $M_Q$ rather than $m_Q$ is relative  ${\cal O}(\Lambda/m_Q)$ and can be neglected.  This Schr\"odinger-like equation implies that  the semi-classical analysis from subsection \ref{A} applies: the spectrum of $  \hat{H}_0$ acting in the space $|\phi\rangle_r$ contains many bound states (with the number scaling with $m_Q$ as $\sqrt{m_Q/\Lambda}$). 

The width, $\Gamma$, quantifies the extent to which $\hat{H}_0^{\rm 2 \, body}$ acts dominantly to describe the states.  From Fermi's Golden rule, $\Gamma$ is given by
\begin{equation}
\begin{split}
\Gamma_j &=2   {}_r\langle\phi_j|\hat{H}_I \hat{G}(E_j)\hat{H}_I |\phi_j \rangle_r \, , \\
{\rm with} \; & \hat{G}(E) \equiv \lim_{\epsilon \rightarrow 0^+}  {\rm Im} \left ( \frac{1}{E -\hat{H}_0^{\rm 2 \, body} + i \epsilon} \right ) \; . \label{Gamma}
\end{split}
\end{equation}  
where $H_I$ was defined in Eq.~(\ref{PQ2}).

 To proceed  we need to  construct the projection operator, $\hat{{\cal P}}^{\rm r}$.  The variable $r$, associated with the separation of the heavy quarks, is a physical quantity and can be expressed in terms of a gauge-invariant operator acting in the relevant space of states.  Consider the gauge invariant operator, $\hat{R}$ defined by 
 \begin{equation}
 \hat{ R}   \equiv  \frac{ \int {\rm d}^3 x  \, {\rm d}^3 y  \, |y|  \,  \hat{Q}^\dagger(\vec{x}) \hat{Q}^\dagger(\vec{x}+\vec{y}) \hat{Q}(\vec{x}+\vec{y}) \hat{Q}(\vec{x}) }{2}  \; .\label{Rdef}
 \end{equation}
$\hat{R}$ measures the distance between the heavy quarks.  Eigenstate  of  $\hat{R}$ has  a fixed value of separation r.    
 
 The eigenstates of $\hat{R}$  are highly degenerate; $\hat{R}$ tells us only about the relative positions of the heavy quarks but nothing about the fast degrees of freedom.  A typical eigenstate of $\hat{R}$ in our space  is denoted 
$|r, \psi_{\rm fast}\rangle$  with 
$\hat{R} |r, \psi_{\rm fast}\rangle = r|r, \psi_{\rm fast}\rangle$; $\psi_{\rm fast}$ represents the state of the fast degrees of freedom. 
 
If the projector $\hat{{\cal P}}^{\rm r}$  exists, then eigenstates of $\hat{R}$ in the projected space are unique.  One can decompose  $\hat{{\cal P}}^{\rm r}$  as
  \begin{equation}
 \begin{split}
  &\hat{{\cal P}}^{\rm r} =\int _0^\infty {\rm d} r \,  |r, \psi_{\rm fast}^{\rm opt}(r) \rangle\langle r, \psi_{\rm fast}^{\rm opt}(r) | \; \; {\rm with} \\
 &\hat{R} |r, \psi_{\rm fast}^{\rm opt}(r)\rangle   = r |r, \psi_{\rm fast}^{\rm opt}(r)\rangle  \, \\
 &\langle r', \psi_{\rm fast}^{\rm opt}(r)| r, \psi_{\rm fast}^{\rm opt}(r)\rangle = \delta(r-r') 
 \end{split}
 \end{equation}
 where last equation fixes the normalization; the label ``opt'' indicates the optimal state  among eigenstates of  $\hat{R}$.    The $r$ dependence in  $\psi_{\rm fast}^{\rm opt}(r)$ makes explicit that the optimal choice depends on $r$.   Determining the optimal state for each $r$ completely fixes the projection operator.

Energetic considerations can be used to determine $\psi_{\rm fast}^{\rm opt}(r)$, in the spirit of the Born-Oppenheimer approximation.  Define a potential operator by subtracting the gauge invariant heavy quark kinetic energy from the leading-order NRQCD Hamiltonian: $\hat{V} \equiv \Hat{H} - \hat{T}^{\rm heavy}$ where $\hat{T}^{\rm heavy} \equiv \int d^3 x Q^\dagger(\vec{x})\left (-\frac{\vec{D}^2}{2 m_Q} \right) Q(\vec{x})$.   Because $\hat{V}$  excludes the only term in $\hat{H}$ that contains derivatives with respect to the positions of the heavy quarks, $\hat{V}$ commutes with $\hat{R}$.  

 Consider  states $| r,  \psi_{\rm fast}(r) \rangle$  that  are  eigenstates of $\hat{R}$  in the physical subspace of the theory.   Matrix elements of $\hat{V}$  defines a potential function of $r$ that depends on $\psi_{\rm fast}$: 
\begin{equation}
V_{\psi_{\rm fast}} ( r)  \delta(r-r') = \langle r',\psi_{\rm fast}|\hat{V}| r, \psi_{\rm fast} \rangle  
\end{equation}
 $ |r, \psi_{\rm fast}^{\rm opt}(r)\rangle$, the optimal choice for $| r, \psi_{\rm fast} \rangle$ minimizes $V_{\psi_{\rm fast} } ( r)$.   The phase of  $ |r, \psi_{\rm fast}^{\rm opt}(r) \rangle$ is arbitrary; for simplicity we take the phase to be real.   As $m_Q \rightarrow \infty$, the motion in the $r$ variable will be slow; the system will be adiabatic and remain in   $\psi_{\rm fast}^{\rm opt}(r)$, the ground state of the fast degrees of freedom.  

The characteristic momenta for the heavy quarks scales as  $\sqrt{m_Q \Lambda}$ while the gauge fields and light quarks  have a  characteristic scale $\Lambda$.  Thus matrix elements of the gauge invariant kinetic energy $\hat{T}^{\rm heavy} = \int d^3 x Q^\dagger(\vec{x})\left (-\frac{\vec{D}^2}{2 \mu} \right) Q(\vec{x})$  are
\begin{equation}
\begin{split}
& \langle r', \psi_{\rm fast}^{\rm opt}(r') |  \hat{T}^{\rm heavy} (\vec{x})|   r, \psi_{\rm fast}^{\rm opt}(r)\rangle \\
 &= \langle r', \psi_{\rm fast}^{\rm opt}(r')|  -\frac{ \partial_r^2}{M_Q}|   r, \psi_{\rm fast}^{\rm opt}(r)\rangle  + {\cal O}\left( \frac{\Lambda^2}{m_Q} \right )\\
&=  - \frac{ \partial_r^2}{M_Q} \delta(r-r')+ {\cal O}\left( \frac{\Lambda^2}{m_Q} \right )
\end{split}
\label{KE} \end{equation}
where again we use $M_Q$ in place of $m_Q$ in the denominator.  
  
There is a subtlety associated with Eq.~(\ref{KE}): 
 Differentiation of  $|  r, \psi_{\rm fast}^{\rm opt}(r) \rangle $  with respect to $r$ acts implicitly on   $\psi_{\rm fast}^{\rm opt}$(r) as well as explicitly on $r$.  To clarify this issue, write  
$| r, \psi_{\rm fast}^{\rm opt}(r) \rangle $ as a tensor product   
$|r, \psi_{\rm fast}^{\rm opt}\rangle =|r\rangle \otimes |\psi_{\rm fast}^{\rm opt}(r)\rangle $ so that
\begin{equation}
\begin{split}
 & \partial_r^2 |   r, \psi_{\rm fast}^{\rm opt}(r)\rangle   = \left (\partial_r^2 |r\rangle \right ) \otimes |\psi_{\rm fast}^{\rm opt}(r)\rangle +\\
&   2\left (\partial_r |r\rangle \right ) \otimes \left ( \partial_r  |\psi_{\rm fast}^{\rm opt}(r)\rangle  \right   )   
 + |r\rangle \otimes \left ( \partial_r^2  |\psi_{\rm fast}^{\rm opt}(r)\rangle  \right   )   \; .
\end{split}
\label{tensorproductform}\end{equation}

The first term in Eq.~(\ref{tensorproductform}) yields the $\frac{\partial^2_r}{M_Q}$ term in the Schr\"odinger equation in Eq.~(\ref{SE}).  The contribution of the second term in Eq.~(\ref{tensorproductform}) to 
$ \langle r', \psi_{\rm fast}^{\rm opt}(r')| \frac{ \partial_r^2}{M_Q}|   r, \psi_{\rm fast}^{\rm opt}(r) \rangle $ 
is strictly zero: $|\psi_{\rm fast}^{\rm opt}(r)\rangle$ is real and normalized so that $\langle \psi_{\rm fast}^{\rm opt}(r)|\partial_r |\psi_{\rm fast}^{\rm opt}(r)\rangle=0$.  Since the fast degrees of freedom vary over scales of order $\Lambda$, the contribution of the third term   in Eq.~(\ref{tensorproductform}) to $ \langle r', \psi_{\rm fast}^{\rm opt}(r') | \frac{ \partial_r^2}{M_Q}|   r, \psi_{\rm fast}^{\rm opt}(r) \rangle $  is ${\cal O}\left (\frac{\Lambda^2}{m_Q} \right )$; it is formally negligible as $m_Q \rightarrow \infty$.

A  generic state 
  in the subspace $r$  can be written as 
\begin{equation} 
|\phi \rangle_{\rm r} = \int_0^\infty dr \, \phi(r)  |   r, \psi_{\rm fast}^{\rm opt}(r) \rangle  \;  ; \label{stateform}
\end{equation}
eigenstates of $\hat{H}_0^{\rm 2 \, body}$ have this form with $\phi(r)$  satisfying
\begin{equation} 
\left(\frac{-\partial^2_r}{M_Q} +V(r) + {\cal O}\left(\frac{\Lambda^2}{m_Q} \right)   \right)\phi_j(r)=E_j \phi_j(r) \; , \label{SE2}
\end{equation} 
 a  Schr\"odinger  equation form anticipated in Eq.~(\ref{SE}).   $V(r)$ corresponds to $\tilde{V}(r)$ from Sect.~\ref{QQ}. 
 
 With these tools established, consider the scaling of the width, $\Gamma$.    From Eq. (\ref{Gamma}), it is clear that the width of an eigenstate of $\hat{H}_0^{cm}$, $j$, with binding energy $B_j$ is given by
  \begin{align}
&\Gamma_j  = \int d r_1 dr_2  \, \phi_j^*(r_1) F(r_1,r_2;-B_j) \phi_j(r_2)  \; \; {\rm with} \label{FA} \\
&F(r_1,r_2; E)  \equiv  \langle r_1, \psi_{\rm fast}^{\rm opt}(r_1) |\hat{H}_I \hat{G}(E) \hat{H}_I| r_2, \psi_{\rm fast}^{\rm opt}(r_2) \rangle \nonumber \end{align}
where $H_I$ is defined in Eq.~(\ref{PQ2}) and $\hat{G}$ in Eq.~(\ref{Gamma}).   From the previous analysis, it is clear that  
\begin{equation} \begin{split}
&\hat{H}_I  | r, \psi_{\rm fast}^{\rm opt}(r)\rangle   =  
 \left ( \hat{T}^{\rm heavy} + \left (\frac{ \partial_r^2}{M_Q} \right ) \right ) | r, \psi_{\rm fast}^{\rm opt}(r) \rangle  \\
&=  \frac{1}{M_Q} \, \left (\partial_r |r\rangle \right ) \otimes \left ( \partial_r  |\psi_{\rm fast}^{\rm opt}(r)\rangle  \right   )   \;  ,
 \label{HI1}
 \end{split}
 \end{equation}
 up to corrections of order $\left(\frac{\Lambda^2}{m_Q} \right)$.  Introducing $X=(r_1+r_2)/2$, $x=r_1-r_2$ and integrating by parts yields
  \begin{widetext}
 \begin{align}
&\Gamma_j  = \frac{1}{m_Q^2}  \int_0^{X_m(B)} \! d X \int_{-\frac{X}{2}}^{ \frac{X}{2}} dx  \, \phi'_j (X-x/2)^*  K(X,x;-B) \phi'_j(X+x/2)  \times \left ( 1+ {\cal O}\left ( \frac{\Lambda}{m_Q} \right ) \right ) \;  , \;  V( X_m(B)) = -(B+ m_\pi) \nonumber\\
&    K(X,x;-B) \equiv \left ( \langle X-x/2| \otimes \partial_{X}  \langle \psi_{\rm fast}^{\rm opt}(X-x/2) |\right ) \hat{G}(E) \left( | X+x/2\rangle \otimes \partial_X |\psi_{\rm fast}^{\rm opt}(X+x/2)\rangle \rangle \right )  \label{F1}
\end{align}
\end{widetext}
where the prime indicates differentiation.  $X_{m}(B)$, the upper bound of the $X$ integral is defined implicitly via $V(x_m(B)) =-(B+m_\pi) $.\footnote{ This follows from the a decomposition of $\hat{G}(E)$ in terms of continuum eigenstates of $\hat{H}_0$ that are outside the space $r$; it is only nonzero for states with energies that overlap $E$.  These eigenstates are necessarily a  tetraquark plus one or more mesons.  The energies of such states are always of greater energy than the energy of the tetraquark state plus mass of the pion.  Thus, the only contributions come from states for which the tetraquark has an energy of less than $E-m_\pi$. On the other hand, the tetraquark itself can be described via a semi-classical wave function and is exponentially suppressed when $ V(r )+m_\pi  > E$ as it is classically forbidden.} Note that $X_m(B)$ is of order $\Lambda^{-1}$ even if $B$ is arbitrarily small..  

 The scaling behavior of  $K(X,x;E) $  can be deduced from general considerations:
\begin{equation}
K(X,x;E) =\sqrt{m_Q \Lambda^3} \, \, \kappa\left(X \Lambda , x \sqrt{m_Q  \Lambda},\frac{E}{\Lambda} \right ) \; ; \label{Kscale}
\end{equation}
$\kappa$ is a dimensionless function of dimensionless variables.\footnote{  $K(X,x;E)$ characterizes the response of the fast degrees of freedom for given values of the $r$; hence the dependence on $E$,  goes as $E/\Lambda$.    In the absence of the heavy quark kinetic energy in the denominator of  $\hat{G}(E)$, the dependence on $x$ would  be a $\delta$-function.  The kinetic energy spreads it out; but since it  has two spatial derivatives  and one factor of  $m_Q$, the dependence must scale as $x \sqrt{m_Q \Lambda}$.   The dependence on $X$ and overall factor of $\sqrt{m_Q \Lambda^3}$ is obtained by requiring the integration $K$ with  respect to $x$ be independent of $m_Q$.}

The scaling  of $\Gamma_j$ depends on the scaling  of $\phi_j(r)$. The system is  semiclassical and  the classically allowed region away from the turning points---the region of dominant contributions---$\phi_j(r)$ is well approximated as  
\begin{align}
&\frac{\phi_j(r)}{\cal N} =\frac{\Lambda^{\frac{3}{4}}\sin\left(\delta_j(r) \right ) }{(E_j-V(r))^{\frac{1}{4}}} \; , \;
\delta_j(r) \! \equiv \! \int_0^r \! dr'\sqrt{M_Q (E_j-V(r'))} \nonumber \\
&{\cal N} =\left (\frac{\Lambda^{\frac{3}{2}}}{2} \int_0^{r_B} \frac{dr'}{\sqrt{E_j-V(r')}} \right )^{-\frac{1}{2}}\label{scphi}\end{align}
where $ V(r_B) =-B$,  ${\cal N}$ is a dimensionless normalization constant, $r_B$ is the turning point.

 $V(r)$ can be written in terms of a dimensionless function, $v$:
$V(r) \equiv \Lambda v(\Lambda r) $.  Thus, for a generic  tetraquark state ${\cal O}(\Lambda)$ below threshold,  ${\cal N}$ is independent of $m_Q$ and
\begin{equation}
\delta_j(r) =\sqrt{\frac{m_Q}{\Lambda}}\tilde{\delta}_j(r \Lambda) \label{deltatilde}
\end{equation}
where $\tilde{\delta}$ is  dimensionless.
Up to corrections of  order $\left (\frac{\Lambda}{m_Q} \right)^{\frac12}$, 
 \begin{widetext}
 \begin{equation} 
{\phi_j'(X \pm \frac x2)}= -{{\cal N} \left (\Lambda^3  m_Q^2 (E_j-\Lambda v(\Lambda X))\right )^{\frac{1}{4}}  }\cos \left (\sqrt{\frac{m_Q}{\Lambda}}\tilde{\delta}_j \left((X \pm \frac x2) \Lambda \right)  \right ) 
\end{equation}
since at large $M_Q$ the  derivative of  $\phi_j'$  is dominated  by the derivative acting on the rapidly varying sine term and, from Eq.~(\ref{Kscale}), $x$ has characteristic values of {\cal O}$\left(\sqrt{\frac{1}{m_Q \Lambda} } \right)$.  Inserting $\phi'$ into Eqs.~(\ref{F1}) and (\ref{Kscale}), yields
\begin{align}
& \Gamma_j  = \frac{|{\cal N}|^2 \Lambda^{\frac32}}{m_Q}  \int_0^{X_m(B) } \! d X  \left(E_j-\Lambda v(\Lambda X))\right )^{\frac{1}{2}}  k_j(X,B)
  \times \left ( 1+ {\cal O}\left ( \frac{\Lambda}{m_Q} \right ) \right ) \\
& k_j(X,B) \equiv \sqrt{m_Q \Lambda^3} \,  \int_{-\frac{X}{2}}^{ \frac{X}{2}} dx  \,  \kappa\left(X \Lambda , x \sqrt{m_Q  \Lambda},-\frac{B}{\Lambda} \right ) \, \cos \left (\sqrt{\frac{m_Q}{\Lambda}}\tilde{\delta}_j \left((X - \frac x2) \Lambda \right)  \right )   \cos \left (\sqrt{\frac{m_Q}{\Lambda}}\tilde{\delta}_j \left((X + \frac x2) \Lambda \right)  \right )  \nonumber
\end{align}
\end{widetext}

The magnitude of the cosines in the integral defining $k_j(X,B)$ are bounded by unity.  This implies that $k_j(X,B)$ is of order  unity or less in a $\Lambda/m_Q$ expansion, which in turn implies
$ \Gamma_j \sim   \frac{\Lambda^2}{m_Q}$  given that ${\cal N}$ is order unity; the scaling of Eq.~(\ref{GammaScal}) is established.  We have thus demonstrated in a model-independent way the principal result of this paper: for sufficiently large heavy quarks masses, there exist parametrically narrow unstable  tetraquarks.  

This argument was explicitly constructed for states with $I = 1$, $J = 0$ and  positive parity.  However, the logic of Subsect.~\ref{OQN} applies here as well and thus the analysis holds in any attractive channel.

\section{Near threshold tetraquarks}

The analysis in Sects.~\ref{Many} and \ref{MCD} was for tetraquarks that were generically of order $\Lambda$ below the  the threshold for separation into two heavy mesons.  With minor modifications it can  be extend to near-threshold tetraquarks:  those below the threshold by a binding energy, $B$, that goes to zero as $M_Q \rightarrow \infty$.  In particular we focus on states with
\begin{equation} 
B \lesssim \frac{\Lambda^{2 - \epsilon}}{M_Q^{1-\epsilon}} \; \; {\rm for \, any \, \epsilon \, with} \; 0<\epsilon< 1  \; .
 \label{nt}
 \end{equation}
 and show that such states must exist as  narrow resonances.  This requires of two things: i) a demonstration that the semi-classical analysis of the Schr\"odinger equation remains valid in this regime showing parametrically near threshold tetraquarks exist in this description and ii) a demonstration that the such states remain  narrow.

In the semi-classical analysis of Sect.~\ref{Many} the number of bound states was scaled with $\sqrt{m_Q/\Lambda}$ if $B$ is fixed as $m_Q \rightarrow \infty$.  Suppose instead it is chosen to decrease with  $m_Q$ ensuring that the states remain  parametrically close to threshold:
\begin{equation}
B=b^2  \, 2 \pi^2 \, \frac{m_\pi^{2-\epsilon}}{m_Q^{1-\epsilon}}  \sim  \frac{\Lambda^{2-\epsilon}}{m_Q^{1-\epsilon}}
\label{Bcond} \end{equation}
where $b$ is a dimensionless numerical constant of order unity and $0<\epsilon\le1$.   Then,  Eq.~(\ref{ineql3}) yields $n_b > b \left({M_Q}/{m_\pi} \right )^{\epsilon/2}$. In the limit of large $M_Q$ this implies that there are a large number of states satisfying Eq.~(\ref{Bcond}); i.e. a large number of near threshold states. If $M_Q$ is sufficient large and $0<\epsilon$,  $n_b \gg 1$ ensuring the validity of Eq.~(\ref{ineql3}).  The precise number of states identified as being ``near threshold'' depends on the arbitrary  choice of  $b$ and $\epsilon$, but at large $M_Q$ there are many such states.

The semi-classical parts of the analysis in Sect.~\ref{MCD} are altered for near threshold  tetraquarks.  The principal differences are the semiclassical density of states is higher than   the generic case so that the tetraquark level spacing scales as  
\begin{equation}
\Delta E \sim \frac{\Lambda^{2-\frac{\epsilon}2}}{m_Q^{1-\frac{\epsilon}2}}
\end{equation}
and the integrals over wave functions are  dominated by the region near the turning point.   Taking this into account gives  $|{\cal N}|^2 \sim \left ( \frac{\Lambda}{m_Q} \right)^{\frac{1}{2}-\frac{\epsilon}{2}}$ so that $
\Gamma \sim \frac{|{\cal N}|^2 \Lambda^2}{m_Q} \sim  \frac{\Lambda^{
\frac{5}{2}-\frac{\epsilon}{2}}}{m_Q^{\frac{3}{2}-\frac{\epsilon}{2}}}$.
Note that  just as in the generic case $\frac{\Gamma}{\Delta E}  \sim \sqrt{\frac{\Lambda}{m_Q}}$ for these near-threshold states; at large $m_Q$ the widths are much smaller than the spacing and the resonances remain parametrically narrow.

\section{Discussion}

 This paper focused on tetraquarks containing  two heavy quarks.  While it has long been known that deeply bound tetraquarks  with these quantum numbers must exist in the heavy quark limit of QCD\cite{cohen2006doubly, manohar1993exotic}, this work showed that multiple narrow tetraquarks must exist in this limit and that some of these will be   close to the threshold.  The conclusion is model-independent.  This process giving rise to the width is pion emission.  However, the analysis did not require any knowledge of the detailed mechanism by which pions were formed and emitted.  Thus the conclusion is robust---provided that $m_Q$ is sufficiently large.

 The result obtained here may be of theoretical interest, but it is not of direct phenomenological relevance.   Tetraquarks with two heavy quarks have yet to be observed experimentally.  Moreover, it seems quite likely that  if tetraquarks of this sort do exist in QCD,  both charm and bottom are too light for the semi-classical analysis used here to be valid.  However, there is a realistic possibility that a fairly large scale separation could exist and some of the analysis could still apply.
 
An obvious question is whether an analysis of this sort can be extended to other  exotic hadrons, either to pentaquarks with two heavy quarks, tetraquarks with  a heavy quark and a heavy antiquark or pentaquarks with a heavy quark and a heavy antiquark.  Apart from the obvious theoretical interest in doing this, there are number of putative exotic hadrons containing a heavy quark and a heavy antiquark---both tetraquarks and pentaquarks\cite{ali2017exotics}.  Such systems are more difficult to control theoretically than the one studied here due to rearrangement effects.   If however, rearrangement processes can be shown to be dynamically suppressed  in the limit of $m_Q \rightarrow \infty$ then an argument similar to the one given here should apply.  This possibility will be explored in future work.

While the analysis in this paper made intricate use of the semi-classical and Born-Oppenheimer  approximations, the underlying physics is rather straightforward: as $m_Q$ increases, there is an increasing large scale separation  between the slow scales associated with the motion of the heavy quarks and the fast degree of freedom.   This forces a decoupling between any process at scale $\Lambda$  (including pion emission) and the slow motion of heavy quarks associated with the potential description of tetraquarks.  It is plausible that this will continue to hold even if $m_Q$ is not large enough to put the system in semi-classical regime that allowed the detailed analysis here but might be large enough to justify a Born-Oppenheimer scale separation.  If that can be shown to be the case, then it may be relevant if sometime in the future doubly-charmed tetraquarks are discovered and the spectrum has an excited state more than $m_\pi$.

 It is plausible that if a way to extend the analysis here to channels where rearrangement is possible, the basic conclusion---that resonances become narrow due to the heaviness  of the quarks---may well hold there.   That would clearly be of phenomenological significance.

The authors thank the United States Department of Energy under grant DE- FG02-93ER-40762for supporting this research.
\bibliography{main}

\end{document}